\documentclass[aps,prd,preprint,preprintnumbers,groupedaddress,showpacs]{revtex4-1}

\newcommand{\e}{\mathrm{e}}
\newcommand{\ep}{\epsilon}
\newcommand{\univ}{\mathrm{univ}}
\newcommand{\vev}[1]{\left\langle #1 \right\rangle}
\newcommand{\vvev}[1]{\left\langle\kern-0.3em\left\langle #1
      \right\rangle\kern-0.3em\right\rangle} 
\newcommand{\Op}{\mathcal{O}}
\newcommand{\K}[1]{K \left( \frac{#1}{\Lambda} \right)}

\newcommand{\kk}[1]{k \left( \frac{#1}{\Lambda} \right)}

\newcommand{\D}[1]{\Delta \left( \frac{#1}{\Lambda} \right)}

\newcommand{\lb}{\left\lbrace}
\newcommand{\rb}{\right\rbrace}

\newcommand{\N}{\mathcal{N}}
\newcommand{\nn}{\nonumber}
\newcommand{\Tr}{\mathrm{Tr}\,}
\newcommand{\Ld}[1]{\frac{\overleftarrow{\delta}}{\delta #1}}
\newcommand{\Rd}[1]{\frac{\overrightarrow{\delta}}{\delta #1}}
\newcommand{\fey}[1]{\hbox{$#1$\kern-0.5em\raise0.3ex\hbox{/}}}

\begin{document}

\preprint{KOBE-TH-15-02}

\title{Equivalence of Wilson Actions}


\author{H. Sonoda}
\email[E-mail: ]{hsonoda@kobe-u.ac.jp}
\affiliation{Physics Department, Kobe University, Kobe 657-8501, Japan}


\date{30 March 2015, revised 5 August 2015}

\begin{abstract}
  We introduce the concept of equivalence among Wilson actions.
  Applying the concept to a real scalar theory on a euclidean space,
  we derive the exact renormalization group transformation of
  K.~G.~Wilson, and give a simple proof of universality of the
  critical exponents at any fixed point of the exact renormalization
  group transformation.  We also show how to reduce the original
  formalism of Wilson to the simplified formalism by J.~Polchinski.
\end{abstract}


\maketitle

\section{Introduction\label{section-introduction}}

The purpose of this paper is to introduce the concept of equivalence
among Wilson actions.  We consider a generic real scalar theory in
$D$-dimensional euclidean space, and denote the Fourier component of
the scalar field with momentum $p$ by $\phi (p)$.

A Wilson action $S [\phi]$ is a real functional of $\phi (p)$.  A
momentum cutoff is incorporated so that the exponentiated action
$\e^{S [\phi]}$ can be integrated with no ultraviolet
divergences.\cite{Wilson:1973jj} An example is given by
\begin{equation}
S [\phi] = - \int_p \frac{p^2}{\K{p}} \frac{1}{2} \phi (p) \phi (-p) +
S_I [\phi] 
\label{S-example}
\end{equation}
where the cutoff function $K(\bar{p})$ is a positive function of
$\bar{p}^2$ that is $1$ at $\bar{p}^2=0$, and decreases toward $0$ rapidly
for $\bar{p}^2 \gg 1$.  The first term of the action can suppress the modes
with momenta higher than $\Lambda$ sufficiently that $\e^{S [\phi]}$
can be integrated over $\phi (p)$ of all momenta.  The second term
consists of local interaction terms.  In the continuum approach
adopted here, correlation functions are defined for $\phi (p)$ of all
momenta, even those above $\Lambda$.

A Wilson action is meant to describe low momentum (energy) physics
accurately, but not physics at or above the cutoff scale.  If two
Wilson actions describe the same low energy physics, we regard them as
equivalent.  It is the purpose of this paper to provide a concrete
definition of equivalence using the continuum approach.

The paper is organized as follows.  In sect.~\ref{section-equivalence}
we introduce modified correlation functions, and then define
equivalence of Wilson actions as the equality of the modified
correlation functions.  Basically, two Wilson actions are equivalent
if their differences can be removed if we give them a massage at their
respective cutoff scales.  We derive two versions of explicit formulas
that relate two equivalent Wilson actions.  The concept of equivalence
is applied in the rest of the paper.

In sect.~\ref{section-ERG} we derive the exact renormalization group
(ERG) transformation of Wilson \cite{Wilson:1973jj} by considering a
particular type of equivalence.  We derive the ERG differential
equation from our equivalence, which amounts to an integral solution
to the differential equation.  In sect.~\ref{section-vs} we discuss
the relation between the original formulation of ERG transformation by
Wilson and the formulation by J.~Polchinski \cite{Polchinski:1983gv}
which is more convenient for perturbation theory.  In the previous
literature only passing remarks have been given on this
relation.\cite{Morris:1994ie}\cite{D'Attanasio:1997he} Our short
discussion of their relation is complete and hopefully illuminating.
Sect.~\ref{section-fixedpoint} prepares us for the discussion of
universality in sect.~\ref{section-univ}.  We generalize the
definition of equivalence so that the exact renormalization group
transformation can have fixed points.  In sect.~\ref{section-univ}, we
assume a fixed point of the ERG transformation, and show that the
critical exponents defined at the fixed point are independent of the
choice of cutoff functions.  This is what we mean by universality.  We
conclude the paper in sect.~\ref{section-conclusion}.

Throughout the paper we work in $D$-dimensional euclidean momentum
space.  We use the following abbreviated notation
\begin{equation}
\int_p \equiv \int \frac{d^D p}{(2 \pi)^D},\quad
\delta (p) \equiv (2 \pi)^D \delta^{(D)} (p)
\end{equation}

\section{Equivalence\label{section-equivalence}}

Given a Wilson action $S[\phi]$, we denote the correlation functions
by
\begin{equation}
\vev{\phi (p_1) \cdots \phi (p_n)}_S \equiv \int [d\phi]\, \phi (p_1)
\cdots \phi (p_n)\, \e^{S[\phi]} \label{vev}
\end{equation}
We consider modifying the correlation functions for high momenta
without touching them for small momenta.  We define modified
correlation functions by
\begin{equation}
\vvev{\phi (p_1) \cdots \phi (p_n)}_S^{K,k} \equiv
\prod_{i=1}^n \frac{1}{K(p_i)} \cdot \vev{\exp \left( - \int_p
    \frac{k(p)}{p^2} \frac{1}{2} \frac{\delta^2}{\delta \phi (p)
      \delta \phi (-p)} \right) \phi (p_1) \cdots \phi (p_n)}_S
\label{vvev}
\end{equation}
where $K (p)$ and $k (p)$ are non-negative functions of $p^2$.  (We
will call them cutoff functions.)  As $p^2 \to 0$, we must find
\begin{equation}
K(p) \longrightarrow 1,\quad
k(p) \longrightarrow 0
\end{equation}
so that the correlation functions are not modified at small momenta.
In addition we constrain $K(p)$ by
\begin{equation}
K(p) \stackrel{p^2 \to \infty}{\longrightarrow} 0
\end{equation}
In other words $K(p)$ is small for $p^2$ larger than the squared
cutoff momentum of the Wilson action.  The fluctuations of $\phi (p)$
with $p$ larger than the cutoff are suppressed, and we enhance their
correlations by the large factor $1/K(p)$ in (\ref{vvev}). The
exponential on the right-hand side of (\ref{vvev}) amounts to mixing a
free scalar with the propagator $-k(p)/p^2$ to the original scalar
field $\phi$.  Since $k(0) = 0$, the free scalar has no dynamics of
its own.

For example, we may take
\begin{equation}
K (p) = \e^{- \frac{p^2}{\Lambda^2}}
\end{equation}
and
\begin{equation}
k (p) = \frac{p^2}{\Lambda^2}\quad\textrm{or}\quad \e^{-
  \frac{p^2}{\Lambda^2}} \left(1 - \e^{- \frac{p^2}{\Lambda^2}}\right)
\end{equation}
where $\Lambda$ is the momentum cutoff $\Lambda$ of the Wilson action
$S$.  

In particular, for $n=2$ and $n=4$, (\ref{vvev}) gives
\begin{eqnarray}
\vvev{\phi (p_1) \phi (p_2)}_S^{K,k} &=& \frac{1}{K(p_1)^2} \left(
    \vev{\phi (p_1) \phi (p_2)}_S - \frac{k(p_1)}{p_1^2} \delta
    (p_1+p_2) \right)\\
\vvev{\phi (p_1) \cdots \phi (p_4)}_S^{K,k} &=& \prod_{i=1}^4
\frac{1}{K(p_i)} \Bigg[ \vev{\phi (p_1) \cdots \phi
      (p_4)}_S\nn\\
&& - \vev{\phi (p_1) \phi (p_2)}_S \frac{k(p_3)}{p_3^2} \delta
(p_3+p_4)
- \vev{\phi (p_3) \phi (p_4)}_S \frac{k(p_1)}{p_1^2} \delta (p_1+p_2)\nn\\
&&\quad + \frac{k(p_1)}{p_1^2} \delta (p_1+p_2) \frac{k(p_3)}{p_3^2} \delta
(p_3+p_4)\nn\\
&& + (\textrm{t-, u-channels}) \quad\Bigg]
\end{eqnarray}
For small momenta, the modified correlation functions (\ref{vvev})
reduce to the ordinary correlation functions (\ref{vev}).

Now, we would like to introduce the concept of equivalence among
Wilson actions.  Let us regard two Wilson actions $S_1, S_2$ as
equivalent if, with an appropriate choice of $K_{1,2}$ and $k_{1,2}$,
their modified correlation functions become identical for any $n$ and
momenta:
\begin{equation}
\vvev{\phi (p_1) \cdots \phi (p_n)}_{S_1}^{K_1, k_1}
= \vvev{\phi (p_1) \cdots \phi (p_n)}_{S_2}^{K_2, k_2} \label{equivalence}
\end{equation}
Since the functions $K_{1,2}, k_{1,2}$ keep the low energy physics intact, $S_1$
and $S_2$ describe the same low energy physics.  In the following we
solve (\ref{equivalence}) to obtain an explicit relation between the
two actions.

We first rewrite (\ref{equivalence}) as
\begin{eqnarray}
&&\vev{\exp \left( - \int_p \frac{k_2 (p)}{p^2} \frac{1}{2}
      \frac{\delta^2}{\delta \phi (p) \delta \phi (-p)} \right) \phi
  (p_1) \cdots \phi (p_n)}_{S_2}\nn\\
&&= \prod_{i=1}^n \frac{K_2 (p_i)}{K_1 (p_i)} \cdot
\vev{\exp \left( - \int_p \frac{k_1 (p)}{p^2} \frac{1}{2}
      \frac{\delta^2}{\delta \phi (p) \delta \phi (-p)} \right) \phi
  (p_1) \cdots \phi (p_n)}_{S_1} 
\end{eqnarray}
Since functional integration by parts gives
\begin{eqnarray}
&&\vev{\exp \left( - \int_p \frac{k (p)}{p^2} \frac{1}{2}
      \frac{\delta^2}{\delta \phi (p) \delta \phi (-p)} \right) \phi
  (p_1) \cdots \phi (p_n)}_{S}\nn\\
&&= \int [d\phi] \e^{S [\phi]}\exp \left( - \int_p \frac{k (p)}{p^2} \frac{1}{2}
      \frac{\delta^2}{\delta \phi (p) \delta \phi (-p)} \right) \phi
  (p_1) \cdots \phi (p_n)\nn\\
&&= \int [d\phi] \phi
  (p_1) \cdots \phi (p_n) \exp \left( - \int_p \frac{k (p)}{p^2} \frac{1}{2}
      \frac{\delta^2}{\delta \phi (p) \delta \phi (-p)} \right)  \e^{S
    [\phi]}
\end{eqnarray}
we obtain
\begin{eqnarray}
&&\int [d\phi] \left(\prod_{i=1}^n \phi (p_i) \right) \cdot \exp \left( - \int_p
    \frac{k_2 (p)}{p^2} \frac{1}{2} \frac{\delta^2}{\delta \phi (p)
      \delta \phi (-p)} \right)  \e^{S_2 [\phi]}\nn\\
&&= \int [d\phi] \left(\prod_{i=1}^n \frac{K_2 (p_i)}{K_1 (p_i)} \phi
    (p_i) \right) \cdot \exp \left( - \int_p
    \frac{k_1 (p)}{p^2} \frac{1}{2} \frac{\delta^2}{\delta \phi (p)
      \delta \phi (-p)} \right)  \e^{S_1 [\phi]}
\end{eqnarray}
This implies that
\begin{equation}
\exp \left( - \int_p
    \frac{k_2 (p)}{p^2} \frac{1}{2} \frac{\delta^2}{\delta \phi (p)
      \delta \phi (-p)} \right)  \e^{S_2 [\phi]}
= \left[\exp \left( - \int_p
    \frac{k_1 (p)}{p^2} \frac{1}{2} \frac{\delta^2}{\delta \phi (p)
      \delta \phi (-p)} \right)  \e^{S_1
  [\phi]}\right]_{\mathrm{subst}}
\end{equation}
where the suffix ``subst'' denotes the substitution of
\begin{equation}
\frac{K_1 (p)}{K_2 (p)} \phi (p) \label{subst}
\end{equation}
for $\phi (p)$ on the right-hand side.  Hence, we obtain an
intermediate result
\begin{equation}
\e^{S_2 [\phi]}
= \exp \left( \int_p \frac{k_2 (p)}{p^2} \frac{1}{2}
    \frac{\delta^2}{\delta \phi (p) \delta \phi (-p)}\right)
\left[\exp \left( - \int_p
    \frac{k_1 (p)}{p^2} \frac{1}{2} \frac{\delta^2}{\delta \phi (p)
      \delta \phi (-p)} \right)  \e^{S_1
  [\phi]}\right]_{\mathrm{subst}} \label{prelim}
\end{equation}
We can rewrite this in two ways.  First, noting that under the
substitution (\ref{subst}), we must substitute
\begin{equation}
\frac{K_2 (p)}{K_1 (p)} \frac{\delta}{\delta \phi (p)}
\end{equation}
for $\frac{\delta}{\delta \phi (p)}$, we obtain the first relation
\begin{equation}
\e^{S_2 [\phi]} = \exp \left[ \int_p \frac{1}{p^2}
\lb k_2 (p) - k_1 (p) \left(\frac{K_2 (p)}{K_1(p)}\right)^2 \rb
\frac{1}{2} \frac{\delta^2}{\delta \phi (p) \delta \phi (-p)} \right]
\exp \left( S_1 \left[ \frac{K_1}{K_2} \phi \right] \right)
\label{1st}
\end{equation}
Note that the two actions are the same for $\phi (p)$ with small $p$,
since the function of $p^2$ in the curly bracket above is negligible
for small $p^2$.  In this sense the two actions differ only by local
terms.

Alternatively, we rewrite (\ref{prelim}) as
\begin{equation}
\e^{S_2 [\phi]}
= \left[ \exp \left[ \int_p \frac{1}{p^2} \lb k_2 (p)
              \left(\frac{K_1 (p)}{K_2 (p)}\right)^2 - k_1 (p) \rb
          \frac{1}{2} \frac{\delta^2}{\delta \phi (p) \delta \phi
            (-p)} \right] \e^{S_1 [\phi]} \right]_{\mathrm{subst}}
\end{equation}
Using the gaussian formula
\begin{eqnarray}
&&\exp \left( \int_p A(p) \frac{1}{2} \frac{\delta^2}{\delta \phi (p)
      \delta \phi (-p)} \right) \exp \left( S[\phi]\right)\nn\\
&&= \int [d\phi'] \exp \left( - \int_p \frac{1}{2 A(p)} \left(\phi'
        (-p) - \phi (-p) \right) \left(
    \phi' (p) - \phi (p)\right) + S \left[\phi'\right] \right)
\end{eqnarray}
(proven in Appendix \ref{appendix-gauss}), we obtain the second relation
\begin{eqnarray}
\e^{S_2 [\phi]}
&=& \int [d\phi'] \exp \Bigg[ - \int_p \frac{p^2}{2 \left( k_2 (p)
          \left(\frac{K_1 (p)}{K_2 (p)}\right)^2 - k_1 (p) \right)}\nn\\
&& \times    \left( \phi' (p) - \frac{K_1 (p)}{K_2 (p)} \phi (p) \right)
\left( \phi' (-p) - \frac{K_1 (p)}{K_2 (p)} \phi (-p) \right) + S_1
[\phi'] \Bigg] \label{2nd}
\end{eqnarray}

We have thus obtained two explicit formulas (\ref{1st}, \ref{2nd})
relating two equivalent Wilson actions.  The remaining sections give
applications of these formulas.  In Appendix \ref{appendix-fermions}
we give corresponding results for a Dirac fermion field.

\section{Exact Renormalization Group \label{section-ERG}}

Let us apply the results of the previous section to derive the exact
renormalization group transformation of K.~G.~Wilson. (sect.~11 of
\cite{Wilson:1973jj}) We choose
\begin{equation}
\lb\begin{array}{c@{~=~}l@{,\quad}c@{~=~}l}
K_1 (p) & K \left( \frac{p}{\Lambda} \right)& k_1 (p) & k \left(
    \frac{p}{\Lambda}\right)\\
K_2 (p) & K \left( \frac{p}{\Lambda \e^{-t}} \right)& k_2 (p) & k \left(
    \frac{p}{\Lambda \e^{-t}}\right)
\end{array}\right.
\end{equation}
so that the two sets of cutoff functions differ only by the choice of
a momentum cutoff.  We demand that the modified correlation functions
(\ref{vvev}) be independent of the momentum cutoff:
\begin{equation}
\vvev{\phi (p_1) \cdots \phi (p_n)}_{S_2}^{K_2, k_2}
= \vvev{\phi (p_1) \cdots \phi (p_n)}_{S_1}^{K_1, k_1}
\label{ERG-equality}
\end{equation}

Now, using the first formula (\ref{1st}), we obtain
\begin{eqnarray}
\e^{S_2 [\phi]} &=& \exp \left[ \int_p \frac{1}{p^2} \lb k
  \left(\frac{p}{\Lambda} \e^t\right) - k
  \left(\frac{p}{\Lambda}\right) \left(
            \frac{K \left(\frac{p}{\Lambda} \e^t\right)}{K \left(
                  \frac{p}{\Lambda} \right)}\right)^2 \rb
    \frac{1}{2} \frac{\delta^2}{\delta \phi (p) \delta \phi (-p)}
\right] \nn\\
&&\qquad\times \exp \left( S_1 \left[ \frac{K
          \left(\frac{p}{\Lambda}\right)}{K
          \left(\frac{p}{\Lambda} \e^t\right)} \phi (p)\right]\right)
\label{ERG-1st-solution}
\end{eqnarray}
By denoting $S_1$ as $S_\Lambda$ and $S_2$ as $S_{\Lambda \e^{-t}}$,
and taking $t$ infinitesimal, we obtain the exact renormalization
group (ERG) differential equation
\begin{eqnarray}
- \Lambda \frac{\partial}{\partial \Lambda} \e^{S_\Lambda [\phi]}
&=& \int_p \left[
\frac{\D{p}}{\K{p}} \phi (p) \frac{\delta}{\delta \phi (p)}\right.\nn\\
&&\,\left.+ \frac{1}{p^2} \left( 2 \frac{\D{p}}{\K{p}} \kk{p} - \Lambda
    \frac{\partial}{\partial \Lambda} \kk{p} \right)
\frac{1}{2} \frac{\delta^2}{\delta \phi (p) \delta \phi (-p)} \right]
\e^{S_\Lambda [\phi]} \label{ERGdiff}
\end{eqnarray}
where we define
\begin{equation}
\D{p} \equiv \Lambda \frac{\partial}{\partial \Lambda} \K{p}
\end{equation}
This amounts to (11.8) of ref.~\cite{Wilson:1973jj}.  For the
particular choice
\begin{equation}
k (p) = K(p) \left(1 - K (p)\right)\label{Polchinski}
\end{equation}
(\ref{ERGdiff}) gets simplified to 
\begin{equation}
- \Lambda \frac{\partial}{\partial \Lambda} \e^{S_\Lambda [\phi]}
= \int_p \left[
\frac{\D{p}}{\K{p}} \phi (p) \frac{\delta}{\delta \phi (p)}
+ \frac{\D{p}}{p^2} 
\frac{1}{2} \frac{\delta^2}{\delta \phi (p) \delta \phi (-p)} \right]
\e^{S_\Lambda [\phi]} \label{ERGpolchinski}
\end{equation}
This was introduced first by J.~Polchinski.\cite{Polchinski:1983gv}

Alternatively, we can use the second formula (\ref{2nd}) which gives
\begin{eqnarray}
\e^{S_{\Lambda \e^{-t}}[\phi]} &=& \int [d\phi'] \exp \left[ -
  \int_p
  \frac{p^2}{ k \left(\frac{p}{\Lambda} \e^t\right)
    \frac{\K{p}^2}{K\left(\frac{p}{\Lambda} \e^t\right)^2} - \kk{p}}
\right.\nn\\
&&\left.\times \frac{1}{2} \left( \phi' (p) - \frac{\K{p}}{K
      \left(\frac{p}{\Lambda} 
      \e^t\right)} \phi (p)\right)
\left( \phi' (-p) - \frac{\K{p}}{K \left(\frac{p}{\Lambda}
      \e^t\right)} \phi (-p)\right) + S_\Lambda [\phi'] \right]
\label{ERG-2nd-solution}
\end{eqnarray}
This is a well known integral solution of the ERG differential
equation (\ref{ERGdiff}).  (This is discussed in details, for example,
in \cite{Igarashi:2009tj}.) Though
mathematically equivalent, our starting point (\ref{ERG-equality}) of
this section is easier to understand than the differential equation
(\ref{ERGdiff}) or its integral solution (\ref{ERG-2nd-solution}).

Earlier in ref.~\cite{Sonoda:2007dj} it was observed that renormalized
correlation functions of QED can be constructed out of correlation
functions of its Wilson action: the modified correlation functions
(\ref{vvev}) coincide with renormalized correlation functions.  Cutoff
independent correlation functions have been also discussed by
O.~J.~Rosten.\cite{Rosten:2010vm}

\section{Polchinski vs. Wilson\label{section-vs}}

As a second application, we consider 
\begin{equation}
\lb\begin{array}{c@{~=~}l@{,\quad}c@{~=~}l}
K_1 (p) & K \left( \frac{p}{\Lambda} \right)& k_1 (p) & k \left(
    \frac{p}{\Lambda}\right)\\
K_2 (p) & K' \left( \frac{p}{\Lambda} \right)& k_2 (p) & k'
\left(\frac{p}{\Lambda}\right) \equiv K' \left(
    \frac{p}{\Lambda}\right) \left( 1 - K'
    \left(\frac{p}{\Lambda}\right)\right) 
\end{array}\right.
\end{equation}
Note that $k_2$ follows Polchinski's convention (\ref{Polchinski}),
which is convenient for perturbative applications.  Given a solution
$S_1$ of the ERG differential equation (\ref{ERGdiff}) with $K_1,
k_1$, we wish to construct an equivalent $S_2$ that solves
(\ref{ERGpolchinski}) with $K_2, k_2$.

As has been shown in the previous section, the modified correlation
functions are independent of $\Lambda$, if the Wilson action satisfies
(\ref{ERGdiff}).  Hence, if $S_1$ and $S_2$ are equivalent at a
particular $\Lambda$, they give the same modified correlation
functions at any $\Lambda$.  In the following let us choose $\Lambda =
1$, and demand $S_1$ and $S_2$ give the same modified correlation
functions.  Using (\ref{1st}) and denoting $S_1$ as $S$ and $S_2$ as
$S'$, we obtain
\begin{equation}
\e^{S' [\phi]} = \exp \left[ \int_p \frac{1}{p^2} \left( k' (p)
  - k (p) \left( \frac{K'
              (p)}{K(p)}\right)^2 \right)\frac{1}{2} 
        \frac{\delta^2}{\delta \phi (p) \delta \phi (-p)} \right]
\exp \left[ S \left[ \frac{K}{K'} \phi \right]\right]
\end{equation}
A particularly simple result follows if we choose $K'$ satisfying
\begin{equation}
k' (p) = k (p) \left( \frac{K' (p)}{K(p)} \right)^2
\end{equation}
This gives
\begin{equation}
K' (p) = \frac{K(p)}{K(p)^2 + k (p)} \cdot K(p)
\end{equation}
With this choice, we obtain
\begin{equation}
S' [\phi] = S \left[ \frac{K^2 + k}{K} \phi\right]
\end{equation}
so that the two equivalent actions are simply related by a linear
change of field variables.

For the particular choice of made in sec.~11 of \cite{Wilson:1973jj}
\begin{equation}
K(p) = \e^{- p^2},\quad k (p) = p^2
\end{equation}
we obtain
\begin{equation}
K' (p) = \frac{1}{1 + p^2 \e^{2 p^2}}
\end{equation}

\section{ERG for fixed points\label{section-fixedpoint}}

We now apply the results of sect.~\ref{section-equivalence} to show
the universality of critical exponents at a fixed point of the ERG
transformation, by which we mean the independence of critical
exponents on the choice of cutoff functions $K, k$.  For the ERG
transformation to have a fixed point, we must change the
transformation given in sect.~\ref{section-ERG} in two
ways:\cite{Wilson:1973jj} first by adopting a dimensionless notation,
and second by introducing an anomalous dimension to the scalar field.
(We elaborate more on these points in Appendix
\ref{appendix-scaling}.)  After these changes, the Wilson action $S_t$
depends on $t$ such that
\begin{equation}
\vvev{\phi (p_1 \e^{\Delta t}) \cdots \phi (p_n \e^{\Delta
    t})}_{S_{t+\Delta t}}^{K,k} =
\e^{\Delta t \cdot n \left( - \frac{D+2}{2} + \gamma \right)}
\vvev{\phi (p_1) \cdots \phi (p_n)}_{S_t}^{K,k}
\label{scaling}
\end{equation}
for the same cutoff functions $K(p), k(p)$ independent of $t$.  This
is the new form of equivalence between $S_t$ and $S_{t+\Delta t}$:
their modified correlation functions are the same up to a scale
transformation.  On the right-hand side, $- \frac{D+2}{2}$ gives the
canonical mass dimension of the field $\phi (p)$ (since this is the
Fourier transform, we obtain $\frac{D-2}{2}-D = - \frac{D+2}{2}$), and
$\gamma$ is the anomalous dimension, taken for simplicity as a
$t$-independent constant.

Let us solve (\ref{scaling}) to obtain $S_{t+\Delta t}$ in terms of
$S_t$.  Following the same line of arguments given in
sect.~\ref{section-equivalence}, we obtain
\begin{equation}
\e^{S_{t+\Delta t} [\phi]}
= \exp \left( \int_p \frac{k(p)}{p^2} \frac{1}{2}
    \frac{\delta^2}{\delta \phi (p)\delta \phi (-p)}\right)
\left[ \exp \left( - \int_p \frac{k(p)}{p^2} \frac{1}{2}
        \frac{\delta^2}{\delta \phi (p) \delta \phi (-p)} \right)
    \e^{S_t [\phi]} \right]_{\mathrm{subst}}
\end{equation}
where ``subst'' stands for the substitution of 
\begin{equation}
\e^{\Delta t \left(\frac{D+2}{2} - \gamma \right)} \frac{K (p)}{K (p
  \e^{\Delta t})}  \phi (p \e^{\Delta t})
\end{equation}
for $\phi (p)$.  Since this substitution implies the substitution of
\begin{equation}
\e^{\Delta t \left( D - \frac{D+2}{2} + \gamma\right)} \frac{K(p
  \e^{\Delta t})}{K(p)} \frac{\delta}{\delta \phi (p \e^{\Delta t})}
\end{equation}
for $\frac{\delta}{\delta \phi (p)}$, we obtain
\begin{equation}
\e^{S_{t+\Delta t} [\phi]} =
\exp \left[ \int_p \frac{1}{p^2} \left( k(p) - k (p \e^{-\Delta t})
        \frac{K (p)^2}{K(p \e^{-\Delta t})^2} \e^{\Delta t \cdot 2
          \gamma} \right) \frac{1}{2} \frac{\delta^2}{\delta \phi (p)
      \delta \phi (-p)} \right] \left[ \e^{S_t [\phi]}
\right]_{\mathrm{subst}}
\end{equation}
Taking $\Delta t$ infinitesimal, we obtain the ERG differential
equation
\begin{eqnarray}
\partial_t \e^{S_t [\phi]}
&=& \int_p \left[ \left(\frac{\Delta (p)}{K(p)} + \frac{D+2}{2} - \gamma
    \right) \phi (p) + p_\mu \frac{\partial \phi (p)}{\partial p_\mu}
\right] \frac{\delta}{\delta \phi (p)} \e^{S_t [\phi]}\nn\\
&& + \int_p \frac{1}{p^2} \left( 2 \frac{\Delta (p)}{K(p)} k (p) + 2
    p^2 \frac{d k(p)}{d p^2} - 2 \gamma k(p) \right) \frac{1}{2}
\frac{\delta^2}{\delta \phi (p) \delta \phi (-p)} \e^{S_t [\phi]}
\label{ERG-dimless}
\end{eqnarray}
For Polchinski's choice
\begin{equation}
k (p) = K(p) \left(1 - K(p)\right)
\end{equation}
this gets simplified to 
\begin{eqnarray}
\partial_t \e^{S_t [\phi]}
&=& \int_p \left[ \left(\frac{\Delta (p)}{K(p)} + \frac{D+2}{2} - \gamma
    \right) \phi (p) + p_\mu \frac{\partial \phi (p)}{\partial p_\mu}
\right] \frac{\delta}{\delta \phi (p)} \e^{S_t [\phi]}\nn\\
&& + \int_p \frac{1}{p^2} \lb\Delta (p) - 2 \gamma K(p) \left(1 -
        K(p) \right) \rb \frac{1}{2}
\frac{\delta^2}{\delta \phi (p) \delta \phi (-p)} \e^{S_t [\phi]}
\end{eqnarray}
which is given in \cite{Igarashi:2009tj}.  For Wilson's choice
\begin{equation}
K(p) = \e^{- p^2},\quad k(p) = p^2
\end{equation}
(\ref{ERG-dimless}) gives
\begin{eqnarray}
\partial_t \e^{S_t [\phi]}
&=& \int_p \left[ \frac{D}{2} \phi (p) + p_\mu \frac{\partial \phi
      (p)}{\partial p_\mu} 
\right] \frac{\delta}{\delta \phi (p)} \e^{S_t [\phi]}\nn\\
&& + \int_p \left(1  - \gamma  + 2 p^2  \right)
\left(\phi (p) \frac{\delta}{\delta \phi (p)} + \frac{\delta^2}{\delta
      \phi (p) \delta \phi (-p)} \right)\e^{S_t [\phi]} 
\end{eqnarray}
which reproduces (11.17) of \cite{Wilson:1973jj} under the identification
\begin{equation}
\frac{d\rho (t)}{dt} = 1 - \gamma
\end{equation}

Now, the anomalous dimension $\gamma$ is chosen for the existence of a
fixed point action $S^*$ that satisfies
\begin{eqnarray}
&&\int_p \left[ \left(\frac{\Delta (p)}{K(p)} + \frac{D+2}{2} - \gamma
    \right) \phi (p) + p_\mu \frac{\partial \phi (p)}{\partial p_\mu}
\right] \frac{\delta}{\delta \phi (p)} \e^{S^* [\phi]}\nn\\
&& + \int_p \frac{1}{p^2} \left( 2 \frac{\Delta (p)}{K(p)} k (p) + 2
    p^2 \frac{d k(p)}{d p^2} - 2 \gamma k(p) \right) \frac{1}{2}
\frac{\delta^2}{\delta \phi (p) \delta \phi (-p)} \e^{S^* [\phi]}
= 0 \label{ERGfp}
\end{eqnarray}
At the fixed
point, the correlation functions obey the scaling law:
\begin{equation}
\vvev{\phi (p_1 \e^t) \cdots \phi
  (p_n \e^t)}^{K,k}_{S^*}
= \e^{t n \left(-\frac{D+2}{2} + \gamma \right)} 
\vvev{\phi (p_1) \cdots \phi (p_n)}_{S^*}^{K,k}
\end{equation}
Only for specific choices of $\gamma$, (\ref{ERGfp}) has an acceptable
solution.  For example, if we assume $S^*$ to be quadratic in $\phi$,
the solution becomes non-local unless $\gamma = 0, -1, -2, \cdots$.
We then obtain
\begin{equation}
S^* = - \frac{1}{2} \int_p \frac{p^{2(1-\gamma)}}{Z K(p)^2 + k (p)
  p^{2(- \gamma)}} \, \phi (p) \phi (-p)
\end{equation}
which gives
\begin{equation}
\vvev{\phi (p) \phi (q)}_{S^*}^{K,k} = \frac{Z}{p^{2(1-\gamma)}}
\delta (p+q)
\end{equation}
where $Z$ is an arbitrary positive constant.  (This is discussed in
Appendix of \cite{Wilson:1973jj}.)

\section{Universality of Critical Exponents\label{section-univ}}

We now discuss universality of critical exponents at an arbitrary
fixed point $S^*$ of the ERG transformation, reviewed in the previous
section.  Universality within the ERG formalism has been shown in
ref.~\cite{Latorre:2000qc}; our discussion below has the merit of
conciseness. (In Appendix \ref{appendix-latorre} we derive those
results of \cite{Latorre:2000qc} relevant to the present paper.)

$S^*$ depends on $K, k$, but we know from
sect.~\ref{section-equivalence} that for any choice of $K, k$ there is
an equivalent action that gives the same modified correlation
functions. (\ref{1st}) gives the equivalent action $S'^*$ for $K', k'$
as
\begin{equation}
\e^{S'^* [\phi]} = \exp \left[ \int_p \frac{1}{p^2} \left( k' (p)
            \frac{K(p)^2}{K'(p)^2} - k (p) \right) \frac{1}{2}
        \frac{\delta^2}{\delta \phi (p) \delta \phi (-p)} \right]
    \exp \left(S^* \left[\frac{K}{K'} \phi\right]\right)
\label{Sprimestar}
\end{equation}
Since the integrand of the exponent vanishes at $p^2 = 0$, $S^*$ and
$S'^*$ differ by local terms.  Since
\begin{equation}
\vvev{\phi (p_1) \cdots \phi (p_n)}^{K', k'}_{S'^*} =
\vvev{\phi (p_1) \cdots \phi (p_n)}^{K,k}_{S^*}
\end{equation}
the anomalous dimension $\gamma$ is independent of the choice of $K,
k$.

Now, the anomalous dimension $\gamma$ is not the only critical
exponent defined at the fixed point $S^*$.  The other exponents appear
as scale dimensions of local composite operators.\cite{Wilson:1973jj}
A composite operator $\Op_y (p)$ with momentum $p$ is a functional of
$\phi$ satisfying
\begin{equation}
\vvev{\Op_y (p \e^t) \phi (p_1 \e^t) \cdots \phi
  (p_n \e^t)}_{S^*}^{K,k}
= \e^{t \lb - y + n \left(-\frac{D+2}{2} + \gamma \right)\rb} 
\vvev{\Op_y (p) \phi (p_1) \cdots \phi (p_n)}_{S^*}^{K,k}
\end{equation}
where the modified correlation functions are defined by
\begin{eqnarray}
&&\vvev{\Op_y (p) \phi (p_1) \cdots \phi (p_n)}_{S^*}^{K,k}\nn\\
&&\equiv \prod_{i=1}^n \frac{1}{K(p_i)} \cdot \vev{\Op_y (p) 
\exp \left( - \int_p \frac{k(p)}{p^2} \frac{1}{2}
    \frac{\delta^2}{\delta \phi (p) \delta \phi (-p)} \right) \phi (p_1)
  \cdots \phi (p_n)}_{S^*}
\end{eqnarray}
and $- y$ is the scale dimension of $\Op_y$.  (The scale dimension
of $\Op_y (x) = \int_p \e^{i p x} \Op_y (p)$ in coordinate space is
$D-y$.)  For the equivalent fixed point action $S'^*$ with $K', k'$,
the corresponding composite operator has the same modified correlation
functions:
\begin{equation}
\vvev{\Op_y (p) \phi (p_1) \cdots \phi (p_n)}_{S^*}^{K,k}
= \vvev{\Op'_y (p) \phi (p_1) \cdots \phi (p_n)}_{{S'}^*}^{K',k'}
\end{equation}
This gives $\Op'_y (p)$ as
\begin{equation}
\Op'_y (p) \e^{S'^* [\phi]} = \exp \left( \int_p \frac{1}{p^2} \left( k' (p)
            \frac{K(p)^2}{K'(p)^2} - k (p) \right) \frac{1}{2}
        \frac{\delta^2}{\delta \phi (p) \delta \phi (-p)} \right)
\left[\Op_y (p) \e^{S^* [\phi]} \right]_{\mathrm{subst}}
\end{equation}
where ``subst'' implies substitution of
\begin{equation}
\frac{K (p)}{K' (p)} \phi (p)
\end{equation}
into $\phi (p)$.  The scale dimension $y$ is thus independent of the
choice of $K, k$.  We conclude that all the critical exponents are
independent of $K, k$.

Before closing this section, we would like to discuss two issues
related to the fixed point action $S^*$.

\subsection{Ambiguity of the fixed point action\label{subsection-ambiguity}}

Given $K, k$, and an appropriate choice of $\gamma$, the fixed point
solution $S^*$ of the ERG differential equation is still not unique.
This is because normalization of the scalar field can be arbitrary.

Given $S^*$, we can construct $S_Z^*$ satisfying
\begin{equation}
\vvev{\phi (p_1) \cdots \phi (p_n)}_{S_Z^*}^{K,k} = Z^{\frac{n}{2}}
\vvev{\phi (p_1) \cdots \phi (p_n)}_{S^*}^{K,k}
\end{equation}
To obtain $S^*_Z$, we set $K_2 = \sqrt{Z} K_1$ and $k_2 = k_1$ in
(\ref{1st}).  We then get
\begin{equation}
\e^{S_Z^* [\phi]}
= \exp \left( - \left(Z-1\right) \int_p \frac{k(p)}{p^2}
    \frac{1}{2} \frac{\delta^2}{\delta \phi (p) \delta \phi (-p)}
\right) \e^{S^* \left[\frac{\phi}{\sqrt{Z}}\right]}
\end{equation}
For example, the $Z$-dependence of the gaussian fixed point ($\gamma =
0$) is given by
\begin{equation}
  S_{G,Z} [\phi] = - \frac{1}{2} \int_p \frac{p^2}{Z K(p)^2 + k (p)}
  \phi (p) \phi (-p) 
\end{equation}

Taking $Z = 1 + 2 \ep$, where $\ep$ is infinitesimal, we obtain
\begin{equation}
S_{1+2\ep}^* [\phi] - S^* [\phi] = \ep \,\N^* [\phi]
\end{equation}
where 
\begin{equation}
\N^* [\phi] \equiv - \int_p \lb \phi (p) \frac{\delta S^*}{\delta
      \phi (p)} + \frac{k(p)}{p^2} \left(
\frac{\delta S^*}{\delta \phi (p)} \frac{\delta S^*}{\delta \phi (-p)}
+ \frac{\delta^2 S^*}{\delta \phi (p) \delta \phi (-p)} \right) \rb
\end{equation}
is a local composite operator satisfying
\begin{equation}
\vvev{\N^* [\phi] \phi (p_1) \cdots \phi (p_n)}_{S^*}^{K,k} 
= n \vvev{\phi (p_1) \cdots \phi (p_n)}_{S^*}^{K,k}
\end{equation}
Obviously, $\N^* [\phi]$, called an equation-of-motion operator in
\cite{Igarashi:2009tj}, has scale dimension $0$.

\subsection{Universal fixed point action?}

We have shown that the modified correlation functions are universal up
to normalization of the scalar field.  We now ask if there is a
universal Wilson action $S_{\mathrm{univ}}^*$ that gives the universal
modified correlation functions as its unmodified correlation
functions:
\begin{equation}
\vev{\phi (p_1) \cdots \phi (p_n)}_{S_{\mathrm{univ}}^*} =
\vvev{\phi (p_1) \cdots \phi (p_n)}_{S^*}^{K,k}
\end{equation}
This implies
\begin{equation}
\e^{S_{\univ}^* [\phi]} = \left[ \exp \left( - \frac{1}{2} \int_p \frac{k(p)}{p^2}
    \frac{\delta^2}{\delta \phi (p) \delta \phi (-p)}
\right) \e^{S^* [\phi]} \right]_{\mathrm{subst}}\label{Sstaruniv}
\end{equation}
where $S^*$ is the fixed point action for $K, k$, and ``subst''
denotes substitution of
\begin{equation}
K(p) \phi (p)
\end{equation}
for $\phi (p)$.  The above result is obtained from (\ref{1st}) by
setting 
\begin{equation}
K_2 = 1,\quad k_2 = 0\label{limit}
\end{equation}
We expect that the right-hand side is independent of $K, k$, i.e.,
$S^*_{\univ}$ has no cutoff.  But we know that the use of a cutoff is
essential for Wilson actions, and there must be something wrong with
$S^*_{\univ}$.

Let us first consider the example of the gaussian fixed point given by
\begin{equation}
S_G [\phi] = - \frac{1}{2} \int_p \frac{p^2}{K(p)^2 + k (p)} \phi (p)
\phi (-p)
\end{equation}
This gives the modified two-point function
\begin{equation}
\vvev{\phi (p) \phi (q)}_{S_G}^{K,k} = \frac{1}{p^2} \delta (p+q)
\end{equation}
(\ref{Sstaruniv}) indeed gives an action free from a cutoff:
\begin{equation}
S_{G, \univ} [\phi] = - \frac{1}{2} \int_p p^2 \phi (p) \phi (-p)
\end{equation}

For interacting theories, though, we expect (\ref{Sstaruniv})
makes no sense.  Let us look at this a little more closely.
As $K_2, k_2$, we choose
\begin{equation}
\lb\begin{array}{c@{~=~}l@{~\equiv~}l}
K_2 (p) & K_{-t} (p) & K (p \e^{-t})\\
k_2 (p) & k_{-t} (p) & k (p \e^{-t})
\end{array}\right.
\end{equation}
In the limit $t \to +\infty$, we obtain (\ref{limit}):
\begin{equation}
\lim_{t\to +\infty} K_{-t} (p) = 1,\quad \lim_{t \to +\infty} k_{-t} (p) = 0
\end{equation}
We then define $S_{-t}^*$ so that
\begin{equation}
\vvev{\phi (p_1) \cdots \phi (p_n)}_{S_{-t}^*}^{K_{-t}, k_{-t}} = \vvev{\phi
  (p_1) \cdots \phi (p_n)}_{S^*}^{K,k}
\end{equation}
$S_{-t}^*$ is related to $S^*$ by the ERG transformation of
sect.~\ref{section-ERG}.  Since the momentum cutoff of $S^*$ is of
order $1$ (we are using the dimensionless convention), that of $S_{-t}^*$
is of order $\e^t$.  Hence, $S_{\univ}^*$ has the infinite momentum
cutoff.  We then expect that the terms of $S_{\univ}^*$ to have
divergent coefficients.

Thus, there is no fixed point action $S^*_{\univ}$ that gives the
correlation functions without modification.

\section{Concluding Remarks\label{section-conclusion}}

We have introduced the concept of equivalence among Wilson actions.
Our equivalence is physically more transparent than the other
formulations of the exact renormalization group via differential
equations or integral formulas.  In particular we have applied our
equivalence to obtain a simple proof of universality of critical
exponents within the ERG formalism.

\appendix
\section{Gaussian Formula\label{appendix-gauss}}

In this appendix we prove the formula
\begin{eqnarray}
&&\exp \left[ \int_p A (p) \frac{1}{2} \frac{\delta^2}{\delta \phi (p)
      \delta \phi (-p)} \right] \exp \left[ S [\phi] \right]\nn\\
&&= \int [d\phi'] \exp \left[ - \int_p \frac{1}{2 A(p)} \phi' (-p) \phi'
    (p) + S \left[\phi+\phi'\right] \right]
\label{appendix-formula}
\end{eqnarray}
Though the following proof requires the positivity of $A(p)$, we
expect the formula to remain valid as long as both hand sides make
sense.  The left-hand side makes sense for any $A (p)$, and the
right-hand side makes sense even if $A (p) < 0$ for some $p$ as long
as convergence of functional integration is provided by the Wilson
action $S$.

It is easy to understand this formula in terms of Feynman graphs.  The
right-hand side implies the coupling of a scalar field $\phi'$ whose
propagator is $A(p)$.  Contracting the pairs of $\phi'$, we obtain the
left-hand side.  More formally, we can prove the equality by comparing
the generating functionals of both hand sides for arbitrary source
$J(p)$.  Let us first compute the generating functional of the
left-hand side:
\begin{equation}
\e^{W_L [J]} \equiv \int [d\phi] \exp \left[ \int_p J(-p) \phi (p) \right]
\exp \left[ \int_p A(p) \frac{1}{2} \frac{\delta^2}{\delta \phi (p)
      \delta \phi (-p)} \right] \exp \left[ S[\phi] \right]
\end{equation}
Integrating this by parts, we obtain
\begin{eqnarray}
\e^{W_L [J]} &=& \int [d\phi]  \exp \left[ S[\phi] \right] 
\exp \left[ \int_p A(p) \frac{1}{2} \frac{\delta^2}{\delta \phi (p)
      \delta \phi (-p)} \right] \exp \left[ \int_p J(-p) \phi (p)
\right]\nn\\
&=& \int [d\phi]  \exp \left[ S[\phi] 
+ \int_p \left( J(-p) \phi (p) + \frac{1}{2} J(-p) A (p) J(p)
\right)\right]
\end{eqnarray}
We next compute the generating functional of the right-hand side:
\begin{eqnarray}
\e^{W_R [J]} &\equiv& \int [d\phi] \exp \left[ \int_p J(-p) \phi (p)
\right]\nn\\
&&\times \int [d\phi'] \exp \left[ - \frac{1}{2} \int_p \frac{1}{A(p)}
    \phi' (p) \phi' (-p) + S [\phi+\phi'] \right]
\end{eqnarray}
We first shift $\phi'$ by $-\phi$, and then shift $\phi$ by $+\phi'$
to obtain
\begin{eqnarray}
\e^{W_R [J]} &=& \int [d\phi] [d\phi']\exp \left[ \int_p J(-p) \left(
        \phi (p) + \phi' (p) \right)\right]\nn\\
&&\times \exp \left[ - \frac{1}{2} \int_p \frac{1}{A(p)} \phi (p) \phi
    (-p) + S [\phi'] \right]\nn\\
&=& \int [d\phi] \exp \left[ \int_p \left( - \frac{1}{2 A(p)} \phi (p)
        \phi (-p) + J(-p) \phi (p) \right)\right]\nn\\
&&\quad\times \int [d\phi'] \exp \left[ \int_p J(-p) \phi' (p) + S
    [\phi'] \right]
\end{eqnarray}
If $A (p)$ is positive, we can perform the gaussian integral over
$\phi$ to obtain
\begin{equation}
\e^{W_R [J]} = \int [d\phi'] \exp \left[ \frac{1}{2} \int_p J(-p) A(p)
    J(p) + \int_p J(-p) \phi' (p) + S[\phi'] \right]
\end{equation}
We thus obtain
\begin{equation}
W_L [J] = W_R [J]
\end{equation}
for arbitrary $J$.  This proves the gaussian formula (\ref{appendix-formula}).

Finally, shifting $\phi'$ by $- \phi$, we rewrite (\ref{appendix-formula})
as
\begin{eqnarray}
&&\exp \left[ \int_p A (p) \frac{1}{2} \frac{\delta^2}{\delta \phi (p)
      \delta \phi (-p)} \right] \exp \left[ S [\phi] \right]\nn\\
&&= \int [d\phi'] \exp \left[ - \int_p \frac{1}{2 A(p)} \left(\phi'
        (-p) -\phi (-p)\right) \left(\phi'
    (p) - \phi (p)\right) + S \left[\phi'\right] \right]
\end{eqnarray}
This is the form used in section \ref{section-equivalence}.

\section{Equivalence of Fermionic Wilson Actions\label{appendix-fermions}}

For a Dirac spinor field $\psi$ and its complex conjugate
$\bar{\psi}$, we define modified correlation functions by
\begin{eqnarray}
    &&\vvev{\psi (p_1) \cdots \psi (p_n) \bar{\psi} (q_n) \cdots \bar{\psi}
      (q_1)}_S^{K,k}
    \equiv \prod_{i=1}^n \frac{1}{K(p_i) K(q_i)} \nn\\
    &&\quad \times
    \vev{\psi (p_1) \cdots \psi (p_n) \exp \left( - \int_p \Ld{\psi (p)}
          \frac{k(p)}{\fey{p}} 
          \Rd{\bar{\psi} (-p)} \right) \bar{\psi} (q_n) \cdots
      \bar{\psi} (q_1)}_S 
\end{eqnarray}
so that
\begin{equation}
\vvev{\psi (p) \bar{\psi} (q)}_S^{K,k} = \frac{1}{K (p)^2} \left[
\vev{\psi (p) \bar{\psi}
  (q)}_S - \frac{k(p)}{\fey{p}} \delta (p+q)\right]
\end{equation}
Two Wilson actions $S_{1,2}$ are equivalent if $K_{1,2}$ and $k_{1,2}$
exist so that
\begin{equation}
\vvev{\psi (p_1) \cdots \psi (p_n) \bar{\psi} (q_n) \cdots \bar{\psi}
  (q_1)}_{S_1}^{K_1, k_1} = 
\vvev{\psi (p_1) \cdots \psi (p_n) \bar{\psi} (q_n) \cdots \bar{\psi}
  (q_1)}_{S_2}^{K_2, k_2} 
\end{equation}

The formula analogous to (\ref{2nd}) is given by
\begin{eqnarray}
\e^{S_2 [\psi,\bar{\psi}]} &=& \int [d\psi' d\bar{\psi}'] \exp \Bigg[
- \int_p \left(\bar{\psi}' (-p) - \frac{K_1 (p)}{K_2 (p)} \bar{\psi} (-p)
\right) \frac{\fey{p}}{k_2 (p) \frac{K_1 (p)^2}{K_2 (p)^2} - k_1
  (p)}\nn\\
&& \qquad \times \left( \psi' (p) - \frac{K_1 (p)}{K_2 (p)} \psi (p) \right) +
  S [\psi', \bar{\psi}'] \Bigg]
\end{eqnarray}
The formula analogous to (\ref{1st}) is somewhat more complicated to
write down.  Denoting
\begin{equation}
A_{ab} (p) \equiv \left(\frac{1}{\fey{p}}\right)_{ab} \left( k_2 (p) -
    k_1 (p) \frac{K_2 (p)^2}{K_1 (p)^2}\right)
\end{equation}
we obtain
\begin{eqnarray}
  \e^{S_2 [\psi,\bar{\psi}]}
  &=& \Tr \left[
    \exp \left( \int_p  \Ld{\psi (p)} A(p) \Rd{\bar{\psi} (-p)}
    \right)
 \exp \left( S_1 \left[ \frac{K_1}{K_2} \psi,
        \frac{K_1}{K_2} \bar{\psi} \right] \right) \right]\\
&\equiv& \sum_{n=0}^\infty \frac{(-)^n}{n!} \int_{p_1,\cdots,p_n}
\prod_{i=1}^n A_{a_i b_i} (p_i) \nn\\
&& \times \Rd{\bar{\psi}_{b_n} (-p_n)} \cdots
\Rd{\bar{\psi}_{b_1} (-p_1)} \exp \left( S_1 \left[\frac{K_1}{K_2}
            \psi, \frac{K_1}{K_2} \bar{\psi}\right] \right)
    \Ld{\psi_{a_1} (p_1)} \cdots \Ld{\psi_{a_n} (p_n)}
\end{eqnarray}
where the spinor indices are summed over.  The exponential implies
contraction of $\psi (p) \bar{\psi} (q)$ by $A (p) \delta (p+q)$.

\section{Derivation of (\ref{scaling})\label{appendix-scaling}}

In this appendix we provide more details behind the new form of
equivalence (\ref{scaling}).  Starting from the original equivalence
(\ref{ERG-equality}), we obtain (\ref{scaling}) in two steps: first by
rescaling dimensionful quantities, and second by introducing an
anomalous dimension of the scalar field.

\subsection{Rescaling}

We first rewrite the equivalence (\ref{ERG-equality}) by rescaling
dimensionful quantities such as momenta and field variables.  Note
that $K_1$ and $K_2$ differ only by a rescaling of momentum
\begin{equation}
K_2 (p \e^{-t}) = K_1 (p)\,.\label{rescaleK}
\end{equation}
Likewise, we have
\begin{equation}
k_2 (p \e^{-t}) = k_1 (p)\,.\label{rescalek}
\end{equation}
We wish to rewrite $S_2$ in such a way that its cutoff functions
become $K_1, k_1$.

For this purpose, we introduce a rescaled field variable
\begin{equation}
\bar{\phi} (p) \equiv \e^{- t \frac{D+2}{2}} \phi (p
\e^{-t}) \label{appendix-rescaling} 
\end{equation}
so that
\begin{equation}
\frac{\delta}{\delta \bar{\phi} (p)} = \e^{- t \frac{D-2}{2}}
\frac{\delta}{\delta \phi (p \e^{-t})}\,. \label{appendix-scaling-delta}
\end{equation}

We then define
\begin{equation}
\bar{S}_2 [\bar{\phi}] \equiv S_2 [\phi]\,.\label{S2bar}
\end{equation}
In other words $\bar{S}_2 [\phi]$ is obtained from $S_2 [\phi]$ by
substituting $\e^{t \frac{D+2}{2}} \phi (p \e^t)$ for $\phi (p)$.  For
example, given
\begin{equation}
S_2 [\phi] = - \frac{1}{2} \int_p \frac{p^2}{K_2 (p)} \phi (p) \phi
(-p)\,,
\end{equation}
we obtain
\begin{eqnarray}
\bar{S}_2 [\phi] &=& - \frac{1}{2} \int_p \frac{p^2}{K_2 (p)} \,\e^{t (D+2)} \phi
(p \e^t) \phi (-p \e^t)\nn\\
&=& - \frac{1}{2} \int_p \frac{p^2}{K_2 (p \e^{-t})} \phi (p) \phi
(-p) = - \frac{1}{2} \int_p \frac{p^2}{K_1 (p)} \phi (p) \phi (-p)\,.
\end{eqnarray}

We rewrite the left-hand side of (\ref{ERG-equality}) as
\begin{eqnarray}
\vvev{\phi (p_1) \cdots \phi (p_n)}_{S_2}^{K_2, k_2}
&=& \e^{n t \frac{D+2}{2}} \vvev{\bar{\phi} (p_1 \e^t) \cdots \bar{\phi}
    (p_n \e^t)}_{S_2}^{K_2, k_2}\nn\\
&=& \e^{n t \frac{D+2}{2}} \prod_{i=1}^n \frac{1}{K_2 (p_i)} \cdot \left\langle
\exp \left( - \int_p \frac{k_2
        (p)}{p^2} \frac{1}{2} \frac{\delta^2}{\delta \phi (p) \delta
        \phi (-p)} \right) \right.\nn\\
&&\left. \quad \times \bar{\phi} (p_1 \e^t) \cdots \bar{\phi}
  (p_n \e^t) \right\rangle_{S_2 [\phi]}\,.
\end{eqnarray}
Using (\ref{rescalek}) and (\ref{appendix-scaling-delta}), we obtain
\begin{eqnarray}
\int_p \frac{k_2 (p)}{p^2} \frac{\delta^2}{\delta \phi (p) \delta \phi
  (-p)}
&=& \e^{- t (D-2)} \int_p \frac{k_2 (p \e^{-t})}{p^2}
\frac{\delta^2}{\delta \phi (p \e^{-t}) \delta \phi (- p
  \e^{-t})}\nn\\
&=& \int_p \frac{k_1 (p)}{p^2} \frac{\delta^2}{\delta \bar{\phi} (p)
  \delta \bar{\phi} (-p)}\,.
\end{eqnarray}
Hence, using (\ref{rescaleK}), we obtain
\begin{eqnarray}
\vvev{\phi (p_1) \cdots \phi (p_n)}_{S_2}^{K_2, k_2}
&=& \e^{n t \frac{D+2}{2}} \prod_{i=1}^n \frac{1}{K_1 (p_i \e^t)} \cdot
\left\langle \exp \left( - \int_p \frac{k_1 (p)}{p^2} \frac{1}{2}
        \frac{\delta^2}{\delta \bar{\phi} (p) \delta \bar{\phi} (-p)}
    \right) \right.\nn\\
&&\quad \times \left.\bar{\phi} (p_1 \e^t) \cdots \bar{\phi} (p_n
\e^t)\right\rangle_{S_2 [\phi]}
\end{eqnarray}
Using (\ref{S2bar}) and rewriting integration variables $\bar{\phi}$
as $\phi$, we obtain
\begin{equation}
\vvev{\phi (p_1) \cdots \phi (p_n)}_{S_2}^{K_2, k_2} = \e^{n t
  \frac{D+2}{2}} \vvev{\phi (p_1 \e^t) \cdots \phi (p_n
  \e^t)}_{\bar{S}_2}^{K_1, k_1}
\end{equation}
Thus, by rescaling, $S_2$ has been converted to $\bar{S}_2$ with the
cutoff functions $K_1, k_1$.

We can now write (\ref{ERG-equality}) as
\begin{equation}
\e^{n t \frac{D+2}{2}} \vvev{\phi (p_1 \e^t) \cdots \phi (p_n
  \e^t)}_{\bar{S}_2}^{K_1, k_1}
= \vvev{\phi (p_1) \cdots \phi (p_n)}_{S_1}^{K_1, k_1}
\end{equation}
Replacing $t$ by $\Delta t$, we obtain
\begin{equation}
\vvev{\phi (p_1 \e^{\Delta t}) \cdots \phi (p_n \e^{\Delta t})}_{\bar{S}_2}^{K_1, k_1} 
= \e^{- \Delta t \cdot n \frac{D+2}{2}} \vvev{\phi (p_1) \cdots \phi
  (p_n)}_{S_1}^{K_1, k_1} \label{zerogamma}
\end{equation}
By writing $S_1$ as $S_t$ and $\bar{S}_2$ as $S_{t+\Delta t}$, we
obtain (\ref{scaling}) for $\gamma = 0$.

\subsection{Anomalous dimension}

Given a Wilson action $S [\phi]$, we can construct an action $S_Z
[\phi]$ whose modified correlation functions differ only by
normalization of the field:
\begin{equation}
\vvev{\phi (p_1) \cdots \phi (p_n)}_{S_Z}^{K,k} = Z^{\frac{n}{2}}
\vvev{\phi (p_1) \cdots \phi (p_n)}_S^{K,k}
\end{equation}
To obtain $S_Z$, we set $K_2 = \sqrt{Z}\,K_1$ and $k_2 = k_1$ in
(\ref{1st}).  We then get
\begin{equation}
\exp \left(S_Z [\phi]\right) = \exp \left( - (Z-1) \int_p
    \frac{k(p)}{p^2} \frac{1}{2} \frac{\delta^2}{\delta \phi (p)
      \delta \phi (-p)} \right) \exp \left( S
    \left[\frac{\phi}{\sqrt{Z}}\right]\right) 
\end{equation}
(We have used the same transformation for the fixed point action in
sect.~\ref{subsection-ambiguity}.)

Given $\bar{S}_2$, we construct $S'_2$ such that
\begin{equation}
\vvev{\phi (p_1) \cdots \phi (p_n)}_{S'_2}^{K_1, k_1} =
\e^{n \gamma \Delta t} \vvev{\phi (p_1) \cdots \phi (p_n)}_{\bar{S}_2}^{K_1,
  k_1}
\end{equation}
where $\gamma$ is an arbitrary constant.  Then, (\ref{zerogamma}) becomes
\begin{equation}
\vvev{\phi (p_1 \e^{\Delta t}) \cdots \phi (p_n \e^{\Delta t})}_{S'_2}^{K_1, k_1} =
\e^{\Delta t \cdot n \left(- \frac{D+2}{2} + \gamma \right)} \vvev{\phi (p_1)
  \cdots \phi (p_n)}_{S_1}^{K_1, k_1}
\end{equation}
This gives (\ref{scaling}), which defines the renormalization group
transformation with an anomalous dimension.

Note that we have introduced an anomalous dimension $\gamma$ by hand.
A particular $\gamma$ must be chosen for the new renormalization group
transformation to have a fixed point.

\section{Relation to the results of Latorre and Morris\label{appendix-latorre}}

In \cite{Latorre:2000qc} Latorre and Morris have shown that the change of a
cutoff function can be compensated by a change of field variables.  We
would like to explain briefly how their result can be reproduced from
the results of the present paper.  

The relation between two equivalent actions $S_1$ (with $K_1, k_1$)
and $S_2$ (with $K_2, k_2$) has been given by (\ref{1st}).  Choosing
\begin{equation}
\lb\begin{array}{c@{~=~}l@{\quad}c@{~=~}l}
K_1 & K,& k_1& k,\\
K_2 & K+\delta K,& k_2& k+\delta k,
\end{array}\right.
\end{equation}
where $\delta K$ and $\delta k$ are infinitesimal, we obtain from
(\ref{1st})
\begin{equation}
\left(S_2 [\phi] - S_1 [\phi]\right) \e^{S_1 [\phi]}
\simeq \int_p \frac{\delta}{\delta \phi (p)}
\left[ \theta (p) \,\e^{S_1 [\phi]} \right]\,,\label{change-of-variables}
\end{equation}
where 
\begin{equation}
\theta (p) \equiv - \frac{\delta K(p)}{K(p)} \phi (p)
+ \frac{1}{p^2} \left(
\frac{1}{2} \delta k (p) - k (p) \frac{\delta K(p)}{K(p)} \right)
    \frac{\delta S_1}{\delta \phi (-p)} \,.\label{theta}
\end{equation}
In deriving (\ref{change-of-variables}), we have taken only the terms
first order in $\delta K$ or $\delta k$, and we have ignored a field
independent constant.  (\ref{change-of-variables}) gives the change of
the action under an infinitesimal change of $\phi (p)$ by $\theta
(p)$.  Upon the choice of the Polchinski convention $k = K(1-K)$,
(\ref{theta}) reduces to
\begin{equation}
\theta (p) = - \frac{\delta K(p)}{2 p^2} \left( \frac{\delta
      S_1}{\delta \phi (-p)} + \frac{2 p^2}{K(p)} \phi (p) \right)
\end{equation}
which reproduces (3.15) of \cite{Latorre:2000qc}.

In addition Latorre and Morris have shown that the ERG transformation
is also a change of variables.  Our ERG differential equation
(\ref{ERG-dimless}) can be rewritten as
\begin{equation}
\partial_t \e^{S_t [\phi]} = \int_p \frac{\delta}{\delta \phi (p)}
\left[ \Psi_t (p) \,\e^{S_t [\phi]} \right]\,,
\end{equation}
where
\begin{eqnarray}
\Psi_t (p) &\equiv& \left(\frac{D+2}{2}-\gamma +  \frac{\Delta
  (p)}{K(p)} \right) \phi (p) + p_\mu 
\frac{\partial \phi (p)}{\partial p_\mu}\nn\\
&&  + \frac{1}{p^2} \left(
    \frac{\Delta (p)}{K(p)} k (p) + p^2 \frac{d k(p)}{dp^2} - \gamma
    k(p) \right) \frac{\delta S_t}{\delta \phi (-p)}\,.
\end{eqnarray}
Thus, $\partial_t S_t$ is the change of the action by an infinitesimal
change of $\phi (p)$ by $\Psi_t (p)$.  Upon the choice $k = K(1-K)$,
the above reduces to
\begin{equation}
\Psi_t (p) = \left(\frac{D+2}{2}-\gamma +  \frac{\Delta
  (p)}{K(p)} \right) \phi (p) + p_\mu 
\frac{\partial \phi (p)}{\partial p_\mu} + \frac{1}{p^2} \left(
    \frac{1}{2} \Delta (p) - \gamma K(p) (1-K(p))\right) \frac{\delta
  S_t}{\delta \phi (-p)}
\end{equation}
which reproduces (2.3) of \cite{Latorre:2000qc} if $\gamma = 0$.

\begin{acknowledgments}
This work was partially supported by the JSPS grant-in-aid \#
25400258.  Preliminary results in this work were presented at ERG2014
held in Lefkada, Greece.  I would like to thank the organizers of
ERG2014 for giving me the opportunity.
\end{acknowledgments}

\bibliography{paperPRD-v2}

\begin{thebibliography}{1}%
\makeatletter
\providecommand \@ifxundefined [1]{%
 \ifx #1\undefined \expandafter \@firstoftwo
 \else \expandafter \@secondoftwo
\fi
}%
\providecommand \@ifnum [1]{%
 \ifnum #1\expandafter \@firstoftwo
 \else \expandafter \@secondoftwo
\fi
}%
\providecommand \enquote [1]{``#1''}%
\providecommand \bibnamefont  [1]{#1}%
\providecommand \bibfnamefont [1]{#1}%
\providecommand \citenamefont [1]{#1}%
\providecommand\href[0]{\@sanitize\@href}%
\providecommand\@href[1]{\endgroup\@@startlink{#1}\endgroup\@@href}%
\providecommand\@@href[1]{#1\@@endlink}%
\providecommand \@sanitize [0]{\begingroup\catcode`\&12\catcode`\#12\relax}%
\@ifxundefined \pdfoutput {\@firstoftwo}{%
 \@ifnum{\z@=\pdfoutput}{\@firstoftwo}{\@secondoftwo}%
}{%
 \providecommand\@@startlink[1]{\leavevmode\special{html:<a href="#1">}}%
 \providecommand\@@endlink[0]{\special{html:</a>}}%
}{%
 \providecommand\@@startlink[1]{%
  \leavevmode
  \pdfstartlink
   attr{/Border[0 0 1 ]/H/I/C[0 1 1]}%
   user{/Subtype/Link/A<</Type/Action/S/URI/URI(#1)>>}%
  \relax
 }%
 \providecommand\@@endlink[0]{\pdfendlink}%
}%
\providecommand \url  [0]{\begingroup\@sanitize \@url }%
\providecommand \@url [1]{\endgroup\@href {#1}{\urlprefix}}%
\providecommand \urlprefix [0]{URL }%
\providecommand \Eprint[0]{\href }%
\@ifxundefined \urlstyle {%
  \providecommand \doi [1]{doi:\discretionary{}{}{}#1}%
}{%
  \providecommand \doi [0]{doi:\discretionary{}{}{}\begingroup
  \urlstyle{rm}\Url }%
}%
\providecommand \doibase [0]{http://dx.doi.org/}%
\providecommand \Doi[1]{\href{\doibase#1}}%
\providecommand \bibAnnote [3]{%
  \BibitemShut{#1}%
  \begin{quotation}\noindent
    \textsc{Key:}\ #2\\\textsc{Annotation:}\ #3%
  \end{quotation}%
}%
\providecommand \bibAnnoteFile [2]{%
  \IfFileExists{#2}{\bibAnnote {#1} {#2} {\input{#2}}}{}%
}%
\providecommand \typeout [0]{\immediate \write \m@ne }%
\providecommand \selectlanguage [0]{\@gobble}%
\providecommand \bibinfo [0]{\@secondoftwo}%
\providecommand \bibfield [0]{\@secondoftwo}%
\providecommand \translation [1]{[#1]}%
\providecommand \BibitemOpen[0]{}%
\providecommand \bibitemStop [0]{}%
\providecommand \bibitemNoStop [0]{.\EOS\space}%
\providecommand \EOS [0]{\spacefactor3000\relax}%
\providecommand \BibitemShut [1]{\csname bibitem#1\endcsname}%
\bibitem{Wilson:1973jj}%
  \BibitemOpen
  \bibfield{author}{%
  \bibinfo {author} {\bibfnamefont{K.}~\bibnamefont{Wilson}}\ and\ \bibinfo
  {author} {\bibfnamefont{J.~B.}\ \bibnamefont{Kogut}},\ }%
  \bibfield{journal}{%
  \Doi{10.1016/0370-1573(74)90023-4}{\bibinfo {journal} {Phys.~Rept.}}\ }%
  \textbf{\bibinfo {volume} {12}},\ \bibinfo {pages} {75} (\bibinfo {year}
  {1974})%
  \bibAnnoteFile{NoStop}{Wilson:1973jj}%
\bibitem{Polchinski:1983gv}%
  \BibitemOpen
  \bibfield{author}{%
  \bibinfo {author} {\bibfnamefont{J.}~\bibnamefont{Polchinski}},\ }%
  \bibfield{journal}{%
  \Doi{10.1016/0550-3213(84)90287-6}{\bibinfo {journal} {Nucl.~Phys.}}\ }%
  \textbf{\bibinfo {volume} {B231}},\ \bibinfo {pages} {269} (\bibinfo {year}
  {1984})%
  \bibAnnoteFile{NoStop}{Polchinski:1983gv}%
\bibitem{Morris:1994ie}%
  \BibitemOpen
  \bibfield{author}{%
  \bibinfo {author} {\bibfnamefont{T.~R.}\ \bibnamefont{Morris}},\ }%
  \bibfield{journal}{%
  \Doi{10.1016/0370-2693(94)90767-6}{\bibinfo {journal} {Phys.~Lett.}}\ }%
  \textbf{\bibinfo {volume} {B329}},\ \bibinfo {pages} {241} (\bibinfo {year}
  {1994}),\ \Eprint{http://arxiv.org/abs/hep-ph/9403340}{arXiv:hep-ph/9403340
  [hep-ph]}%
  \bibAnnoteFile{NoStop}{Morris:1994ie}%
\bibitem{D'Attanasio:1997he}%
  \BibitemOpen
  \bibfield{author}{%
  \bibinfo {author} {\bibfnamefont{M.}~\bibnamefont{D'Attanasio}}\ and\
  \bibinfo {author} {\bibfnamefont{T.~R.}\ \bibnamefont{Morris}},\ }%
  \bibfield{journal}{%
  \Doi{10.1016/S0370-2693(97)00866-6}{\bibinfo {journal} {Phys.~Lett.}}\ }%
  \textbf{\bibinfo {volume} {B409}},\ \bibinfo {pages} {363} (\bibinfo {year}
  {1997}),\ \Eprint{http://arxiv.org/abs/hep-th/9704094}{arXiv:hep-th/9704094
  [hep-th]}%
  \bibAnnoteFile{NoStop}{D'Attanasio:1997he}%
\bibitem{Igarashi:2009tj}%
  \BibitemOpen
  \bibfield{author}{%
  \bibinfo {author} {\bibfnamefont{Y.}~\bibnamefont{Igarashi}}, \bibinfo
  {author} {\bibfnamefont{K.}~\bibnamefont{Itoh}},\ and\ \bibinfo {author}
  {\bibfnamefont{H.}~\bibnamefont{Sonoda}},\ }%
  \bibfield{journal}{%
  \Doi{10.1143/PTPS.181.1}{\bibinfo {journal} {Prog.~Theor.~Phys.~Suppl.}}\ }%
  \textbf{\bibinfo {volume} {181}},\ \bibinfo {pages} {1} (\bibinfo {year}
  {2010}),\ \Eprint{http://arxiv.org/abs/0909.0327}{arXiv:0909.0327 [hep-th]}%
  \bibAnnoteFile{NoStop}{Igarashi:2009tj}%
\bibitem{Sonoda:2007dj}%
  \BibitemOpen
  \bibfield{author}{%
  \bibinfo {author} {\bibfnamefont{H.}~\bibnamefont{Sonoda}},\ }%
  \bibfield{journal}{%
  \Doi{10.1088/1751-8113/40/31/034}{\bibinfo {journal} {J.~Phys.}}\ }%
  \textbf{\bibinfo {volume} {A40}},\ \bibinfo {pages} {9675} (\bibinfo {year}
  {2007}),\ \Eprint{http://arxiv.org/abs/hep-th/0703167}{arXiv:hep-th/0703167
  [HEP-TH]}%
  \bibAnnoteFile{NoStop}{Sonoda:2007dj}%
\bibitem{Rosten:2010vm}%
  \BibitemOpen
  \bibfield{author}{%
  \bibinfo {author} {\bibfnamefont{O.~J.}\ \bibnamefont{Rosten}},\ }%
  \bibfield{journal}{%
  \Doi{10.1016/j.physrep.2011.12.003}{\bibinfo {journal} {Phys.~Rept.}}\ }%
  \textbf{\bibinfo {volume} {511}},\ \bibinfo {pages} {177} (\bibinfo {year}
  {2012}),\ \Eprint{http://arxiv.org/abs/1003.1366}{arXiv:1003.1366 [hep-th]}%
  \bibAnnoteFile{NoStop}{Rosten:2010vm}%
\bibitem{Latorre:2000qc}%
  \BibitemOpen
  \bibfield{author}{%
  \bibinfo {author} {\bibfnamefont{J.~I.}\ \bibnamefont{Latorre}}\ and\
  \bibinfo {author} {\bibfnamefont{T.~R.}\ \bibnamefont{Morris}},\ }%
  \bibfield{journal}{%
  \Doi{10.1088/1126-6708/2000/11/004}{\bibinfo {journal} {JHEP}}\ }%
  \textbf{\bibinfo {volume} {0011}},\ \bibinfo {pages} {004} (\bibinfo {year}
  {2000}),\ \Eprint{http://arxiv.org/abs/hep-th/0008123}{arXiv:hep-th/0008123
  [hep-th]}%
  \bibAnnoteFile{NoStop}{Latorre:2000qc}%
\end{thebibliography}%

\end{document}